\def\pmb#1{\setbox0=\hbox{#1}%
   \kern-.025em\copy0\kern-\wd0
   \kern.05em\copy0\kern-\wd0
   \kern-0.025em\raise.0433em\box0}
\def\gta{\mathrel{{\lower 3pt\hbox{$\mathchar"218$}}\hskip-8pt
   \raise 2pt\hbox{$\mathchar"13E$}}}
\def\lta{\mathrel{{\lower 3pt\hbox{$\mathchar"218$}}\hskip-8pt
   \raise 2pt\hbox{$\mathchar"13C$}}}
\def\half{{\scriptstyle{1\over2}}}
\def\dagg{\phantom{\dagger}}            

\documentclass[twocolumn,showpacs,prl]{revtex4}

\usepackage{graphicx}
\usepackage{dcolumn}
\usepackage{bm}

\begin{document}

\title{A unified theory for the cuprates, iron-based and similar 
superconducting systems: \\ 
non-Fermi-liquid to Fermi-liquid crossover, low-energy and waterfall
anomalies }

\author{J. Ashkenazi}
 \email{jashkenazi@miami.edu}
\affiliation{%
Physics Department, University of Miami, P.O. Box 248046, Coral
Gables, FL 33124, U.S.A.\\
}%

\date{\today}

\begin{abstract}
A unified theory is outlined for the cuprates, Fe-based, and related
superconductors. Their low-energy excitations are approached in terms of
auxiliary particles representing combinations of atomic-like electron
configurations, and the introduction of a Lagrange Bose field enables
their treatment as bosons or fermions. This theory correctly describes
the observed phase diagram of the cuprates, including the
non-Fermi-liquid to FL crossover in the normal state, the existence of
Fermi arcs below $T^*$ and of ``marginal-FL'' behavior above it. The
anomalous behavior of numerous physical quantities is accounted for,
including kink- and waterfall-like spectral features, the drop in the
scattering rates below $T^*$ and more radically below $T_c$, and an
effective increase in the density of carriers with $T$ and $\omega$,
reflected in transport, optical and other properties. Also is explained
the correspondence between $T_c$, the resonance-mode energy, and the
increase in the gap below $T_c$. 
\end{abstract}

\pacs{71.10.Hf, 71.10.Li, 71.10.Pm, 74.20.-z, 74.20.Mn, 74.20.Rp,
74.25.Dw, 74.25.Jb, 74.72.-h} 

   \keywords{superconductivity, cuprates, iron, pnictides, auxiliary
particles} 
\maketitle


High-$T_c$ superconductivity (SC) has been in the forefront of
condensed-matter physics research since its discovery in the cuprates
over 22 years ago. This system is characterized by anomalous behavior of
many of its physical properties \cite{Takagi, Hwang, Fisher, Tanaka,
Puchkov, Zasadzinski, Luo, Tanner1, Boeb} which led to the suggestion of
non-Fermi-liquid (non-FL) models \cite{Anderson, Varma}. The recent
discovery of Fe-based high-$T_c$ SCs (referred to below as FeSCs)
provides a new test case for high-$T_c$ theories, especially in view of
the striking similarity of their anomalous properties to those of the
cuprates (see Ref.~\cite{AshkHam}). 

Recently \cite{AshkHam}, a unified theory for the cuprates and the FeSCs
has been derived by the author on the basis of common features in their
electronic structures, including quasi-two-dimensionality, and the
large-$U$ nature of the low-energy electron orbitals. A Hamiltonian,
${\cal H}$, has been written down \cite{AshkHam}, where such electrons 
are approached in terms of auxiliary particles \cite{Barnes},
representing combinations of atomic-like electron configurations, and a
Lagrange field maintaining a constraint imposed on them. 

The auxiliary particles consist of \cite{AshkHam}: ({\it i}) boson
``svivons'' which represent configurations of the undoped ``parent''
compounds, and their condensation results in static or dynamical
inhomogeneities, and consequently in a resonance mode (RM); ({\it ii})
fermion ``quasi-electrons'' (QEs) which represent configurations with an
excess of an electron or a hole over those of the parent compounds, and
their dynamics determines charge transport; ({\it iii}) boson
``lagrons'' of the constraint Lagrange field which represents an
effective fluctuating potential, enabling the treatment of svivons and
QEs as bosons and fermions. 

Within a one-orbital model for the hole-doped cuprates \cite{Anderson} a
svivon state of spin $\sigma$ ($= \uparrow$ or $\downarrow$) at site $i$
is created by $s_{i\sigma}^{\dagger}$, and a QE state is created by
$q_{i}^{\dagger} \equiv h_{i}^{\dagg}$, where $h_{i}^{\dagger}$ creates
a Zhang-Rice singlet. ${\cal H}$ \cite{AshkHam} is expressed in terms of
the auxiliary-particle operators and parameters determined in
first-principles calculations by few groups (see Ref.~\cite{AshkHam}).
The set of parameters used is: $t = 0.43\;$eV, $t^{\prime} = -0.07\;$eV,
$t^{\prime\prime} = 0.03\;$eV, and $J = 0.1\;$eV. Unlike in RVB models
\cite{Anderson}, the condensation-induced inhomogeneities result here in
a multi-component scenario. 

The svivon and QE Green's functions (GF) consist of ``normal'' and
(within their condensates) ``anomalous'' terms, based on $\langle
s_{i\sigma}^{\dagg} s_{j\sigma}^{\dagger} \rangle$ and $\langle
q_{i}^{\dagg} q_{j}^{\dagger} \rangle$, or $\langle
s_{i\uparrow}^{\dagg} s_{j\downarrow}^{\dagg} \rangle$ and $\langle
q_{i}^{\dagg} q_{j}^{\dagg} \rangle$, respectively. The creation
operator of an electron is expressed as \cite{AshkHam}:
$d_{i\sigma}^{\dagger} \cong s_{i\sigma}^{\dagger} q_{i}^{\dagger}$,
and, thus, the normal (anomalous) electron GF could be diagrammatically
presented, at {\it the zeroth-order}, as a convolution of bubble
diagrams of normal (anomalous) QE and svivon GF. Let us denote the
matrix of these approximate electron GF by $\underline{\cal G}_0$
($\underline{\cal F}_0$); it is diagonal in the ${\bf k}$
representation. ${\cal H}$ \cite{AshkHam} introduces to $\underline{\cal
G}_0$ ($\underline{\cal F}_0$) self-energy corrections due to multiple
scattering of QE-svivon pairs. 

For un-paired QEs, the corrections to $\underline{\cal G}_0$ can be
expressed as a sum of a geometrical series in powers of $\underline{\cal
G}_0 \underline{t}$ (where the matrix $\underline{t}$ consists of the
transfer and intrasite terms in ${\cal H}$), yielding $(\underline{1} -
\underline{\cal G}_0 \underline{t})^{-1}$. Using Dyson's equation, the
normal electron GF matrix is expressed as: 
\begin{equation} 
\underline{\cal G} = \underline{\cal G}_0 (\underline{1} -
2\underline{\cal G}_0 \underline{t})^{-1} (\underline{1} -
\underline{\cal G}_0 \underline{t}). 
\label{eq6} 
\end{equation}
Thus, there are two types of poles in $\underline{\cal G}$; while
$\underline{\cal G}_0$ contributes a non-FL distribution of convoluted
QE-svivon poles, $(\underline{1} - 2\underline{\cal G}_0
\underline{t})^{-1}$ introduces FL-like electron poles. Indeed, the
phase diagram of the cuprates \cite{Honma} indicates a non-FL to FL
crossover between the underdoped and the overdoped normal-state regimes
\cite{Takagi, Boeb}. 

\begin{figure*}[t] 
\begin{center}
\includegraphics[width=3.25in]{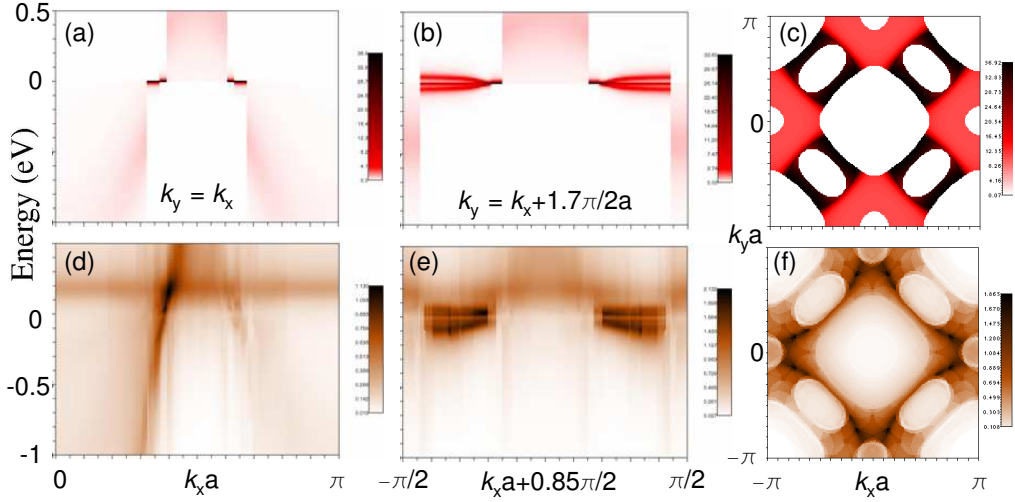}
\end{center}
\caption{The QE (a-c) and electron (d-f) spectral functions in optimally
hole-doped cuprates along a ``line of nodes'' (a,d), along a parallel
line (b,e) shifted by $0.425\pi({\hat y}-{\hat x})$, and on the ``Fermi
surface'' (c,f), determined at $\omega = -0.1 k_{_{\rm B}}T$.} 
\label{fig1}
\end{figure*}

The auxiliary-particle spectra are evaluated through a self-consistent
second-order diagrammatic expansion \cite{AshkHam}. The lagrons are soft
at wave vectors corresponding to major spin and charge fluctuations
(where an inhomogeneous enforcement of the constraint is essential).
Consequently, their energy band $\omega^{\lambda}({\bf q})$ is
characterized \cite{AshkHam} by four V-shape minima $\ll k_{_{\rm B}}T$
at ${\bf Q}_m = {\bf Q} + \delta {\bf q}_m$, where ${\bf Q} =
{\pi \over a}({\hat x}+{\hat y})$ is the parent-compound's
antiferromagnetic (AF) wave vector, and $\delta {\bf q}_m = \pm \delta q
{\hat x} \ \text{or} \ \pm \delta q {\hat y}$ are modulations
corresponding to striped structures (typically $\delta q \cong \pi/4a$).
Degenerate svivon condensates are obtained \cite{AshkHam}, with energy
minima at $\pm {\bf Q}_m/2$, and in the absence of symmetry-breaking
long-range order, the system is in a combination of these states. 

The QE spectrum is evaluated treating the fluctuations between the
combined svivon condensates adiabatically. By introducing appropriate
phase factors to the QE and svivon states, their ${\bf k}$ values are
shifted by ${\bf Q}/2$ or $-{\bf Q}/2$, resulting in a correspondence
between the QE and electron Brillouin zones (BZs). The evaluated QE
spectral functions for hole-doped cuprates, at an optimal stoichiometry
(corresponding to $n^s \equiv \langle s_{i\sigma}^{\dagger}
s_{i\sigma}^{\dagg} \rangle = 0.42$), within the ``hump phase''
\cite{Honma} ($k_{_{\rm B}}T = 0.01\;$eV), are presented in
Fig.~\ref{fig1}(a-c). The QE spectrum has a ``bare'' and a ``dressed''
part, due to coupling to svivon fluctuations and lagrons \cite{AshkHam}.
The bare part is independent of $t$, and determined by \cite{AshkHam}
${\bar t}^{\prime} = t^{\prime} + 4n^sJ$ and ${\bar t}^{\prime\prime} =
t^{\prime\prime} + 2n^sJ$, resulting in {\it equivalent} ``main'' and
``shadow'' QE bands. 

The electron spectral functions, obtained on the basis of
Eq.~(\ref{eq6}), and the QE and svivon \cite{AshkHam} spectra in this
phase, are presented in Fig.~\ref{fig1}(d-f). The high-weight of the
svivon poles around their (averaged) minimal energies results in
band-like features in the electron spectrum which follow the QE band.
The FL-like electron poles in Eq.~(\ref{eq6}) break the equivalence
between the main and shadow bands, in agreement with experiment
\cite{Campuzano, Chang}. 

Coupling of QEs to svivon fluctuations \cite{AshkHam} depends on $t$,
and introduces shifts in their energies towards zero. This results in BZ
areas of low-energy ``flat'' QE bands and apparent discontinuities to
areas of higher energies (see Fig.~\ref{fig1}(a-b)). These features are
modified in the electron spectrum (see Fig.~\ref{fig1}(d-e)) to the
observed \cite{Shen1} low-energy kinks and waterfall-like features above
them. 

The low-energy QE and svivon spectra are strongly renormalized by their
coupling to lagrons (through coefficients $\gamma({\bf q})$). The
resulting QE self-energy corrections (to $\epsilon^q_0({\bf k})$) are
expressed, for un-paired QEs, as: 
\begin{eqnarray}
\Sigma^q({\bf k},z) &\cong& {1 \over N} \sum_{\bf q} |\gamma({\bf q})|^2
\int d\omega^q A^q({\bf k}-{\bf q}, \omega^q) \nonumber \\ &\ &\times
\bigg[ {b_{_T}(\omega^{\lambda}({\bf q})) + f_{_T}(-\omega^q) \over z -
\omega^{\lambda}({\bf q}) - \omega^q} \label{eq7} \\ &\ &\ \ +
{b_{_T}(\omega^{\lambda}({\bf q})) + f_{_T}(\omega^q) \over z +
\omega^{\lambda}({\bf q}) - \omega^q} \bigg],  \ \text{where}
\nonumber \\
A^q({\bf k}, \omega) &=& {\Gamma^q({\bf k}, \omega)/2\pi \over [\omega -
\epsilon^q_0({\bf k}) - \Re \Sigma^q({\bf k},\omega)]^2 +
[\half\Gamma^q({\bf k}, \omega)]^2} \ \ \
\label{eq8} 
\end{eqnarray}
are the QE spectral functions, $\Gamma^q({\bf k}, \omega)$ are their
scattering rates, and $b_{_T}$ ($f_{_T}$) is the Bose (Fermi)
distribution function. The low-energy $\Sigma^q$ is dominated by the
contribution of lagrons around their minima $0 < \omega^{\lambda}({\bf
Q}_m) \ll k_{_{\rm B}}T$ (referred to below as ``${\bf Q}_m$ lagrons''),
and of those of energy $\sim\omega^{\lambda}({\bf Q})$ at their
``extended saddle point'' around ${\bf Q}$ \cite{AshkHam} (referred to
below as ``${\bf Q}$-ESP lagrons''). 

\begin{figure}[t] 
\begin{center}
\includegraphics[width=3.25in]{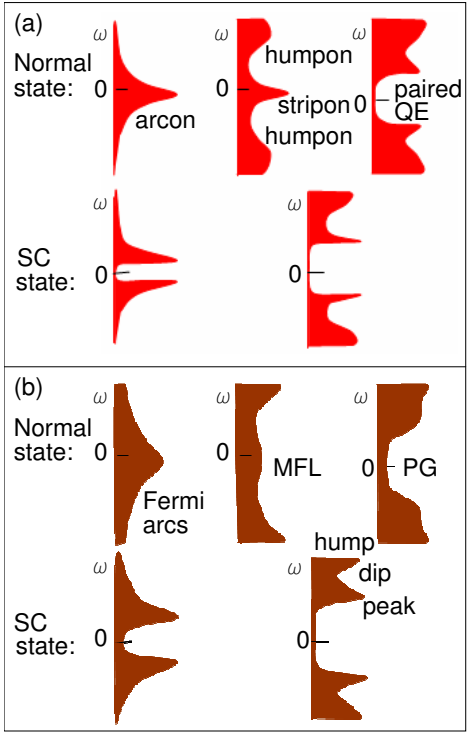}
\end{center}
\caption{Low-energy QE (a) and electron (b) spectral functions in the
normal and SC states, close to the nodal (arcon, Fermi arcs) and
antinodal (humpon, stripon, MFL, paired QE, PG, peak, dip, hump) points
in hole-doped cuprates.} 
\label{fig2}
\end{figure}

As can be seen in Fig.~\ref{fig1}, low-energy QEs around the ``nodal
points'' $\pm{\pi \over 2} ({\hat x}\pm{\hat y})$ are not coupled to
other low-energy QEs through ${\bf Q}_m$ lagrons, and the spectral
functions of such QEs (referred to as ``arcons'') are characterized by a
simple peak, as is sketched in Fig.~\ref{fig2}(a). On the other hand,
low-energy QEs around the ``antinodal points'' $\pm\pi{\hat x}$ and
$\pm\pi{\hat y}$ are coupled to other low-energy QEs through ${\bf Q}_m$
lagrons, resulting in the splitting of the spectral-function peaks of
such QEs (due to the inhomogeneities) into positive- and negative-energy
``humpons''. For dynamical inhomogeneities, a ``stripon'' peak remains
between the humpons, as is sketched in Fig.~\ref{fig2}(a). 

The humpon peaks are rather wide, while the width of the stripon peaks
(see below), as well as their position relative to zero (determined by
the effective QE chemical potential $\mu - \lambda$ \cite{AshkHam}),
scale with $T$. The ${\bf Q}$-ESP lagrons maintain an energy separation
$>\omega^{\lambda}({\bf Q})$ between the stripon and the two humpons.
Analogous expressions to Eq.~(\ref{eq7}) are obtained for the normal and
anomalous svivon self-energy corrections due to their coupling to
lagrons. The effect of the ${\bf Q}$-ESP lagrons there is the
stabilization of high-weight svivon states below
$\sim\omega^{\lambda}({\bf Q})$. Thus \cite{AshkHam}, the RM energy
($E_{_{\rm RM}}$) is $\sim\omega^{\lambda}({\bf Q})$. 

Eq.~(\ref{eq7}) yields a peculiar behavior for $\Gamma^q({\bf k},\omega)
\equiv 2 \Im \Sigma^q({\bf k}, \omega-i0^+)$. In the low $\omega/T$
limit the dominant contribution to $\Gamma^q$ (for stripons) is due to
${\bf Q}_m$ lagrons; the integration of $b_{_T}(\omega^{\lambda})$
yields for them a factor $\propto T^2$ which is multiplied by a factor
$\propto 1/T$ due to the coupled-stripon spectral functions (see
Eq.~(\ref{eq8})), resulting in $\Gamma^q({\bf k},\omega) \propto T$. In
the high $\omega/T$ limit, the dominant contribution to $\Gamma^q$ is
due to the continuum of lagrons \cite{AshkHam} where
$b_{_T}(\omega^{\lambda}) + f_{_T}(\omega^{\lambda} \pm \omega)$ is
either $\sim 1$ or $\sim 0$; the ${\bf q}$ summation in Eq.~(\ref{eq7})
then turns into an integration on the normalized $A^q$, resulting in
$\Gamma^q({\bf k},\omega) \propto \omega$. 

A similar behavior is obtained in this phase also for the svivon
scattering rates $\Gamma^s$, through analogous expressions to
Eq.~(\ref{eq7}). The electron scattering rates are determined through
$\underline{\cal G}$ (see Eq.~(\ref{eq6})). Let $\epsilon^q({\bf k}_1) +
\half i\Gamma^q({\bf k}_1)$ and $\epsilon^s({\bf k}_2) + \half
i\Gamma^s({\bf k}_2)$ be poles of QE and svivon GF, respectively; the
${\bf k} = {\bf k}_1 + {\bf k}_2$ element of $\underline{\cal G}_0$
contributes to  $\underline{\cal G}$ a pole at $\epsilon^q({\bf k}_1)
\pm \epsilon^s({\bf k}_2) + \half i[\Gamma^q({\bf k}_1) + |\Gamma^s({\bf
k}_2)|]$, with a weight scaling with $b_{_T}(\epsilon^s({\bf k}_2) \pm
\half i\Gamma^s({\bf k}_2)) + f_{_T}(\mp(\epsilon^q({\bf k}_1) + \half
i\Gamma^q({\bf k}_1)))$. (Different poles correspond to different
arcons, stripons, humpons, as well as positive and negative svivon
energies \cite{AshkHam}). Thus, if the role of such convoluted poles is
dominant in $\underline{\cal G}$, the electron scattering rates have a
similar dependence on $\omega$ and $T$ as $\Gamma^q$, and this
corresponds to a marginal-Fermi-liquid (MFL) phase \cite{Varma}. 

The signature of the low-energy QE spectral peaks on the electron
spectrum is sketched in Fig.~\ref{fig2}(b). The convolution with svivons
broadens the trace of the stripon peak in the MFL phase; however, since
many quantities are expressed in terms of QE and svivon contributions,
they can be sensitive to its small width. The $T$ and $\omega$
dependencies of quantities determined by the QEs are then characterized
by three energy scales, and crossovers between them; at the lowest
energies only stripons and arcons contribute, while the humpons play a
role in higher energies, and the almost detached higher-energy QEs (see
Fig.~\ref{fig1}(a-b)) have an effect only at the highest energies. 

Such a behavior is observed in the scaling of the $T$-dependence of
numerous quantities \cite{Luo} in the humpon energy. Also, an effective
increase in the carriers density with $T$ and $\omega$ is observed
through the Hall constant \cite{Hwang} and the optical conductivity
\cite{Tanner1}. The low-$T$ dependence of the thermoelectric power has
been analyzed in terms of a ``narrow-band model'' \cite{Fisher}
(consistently with the effect of stripons and arcons), and the growing
contribution of humpons at higher $T$ results in a peculiar behavior
\cite{Tanaka} used to determine the doping level. Non-FL behavior,
associated with the QE Fermi surface (see Fig.~\ref{fig1}(c)), appears
\cite{Chang} as being due to hole and electron pockets around the nodal
and antinodal points, respectively. 

QE pairing is approached adapting the Migdal-Eliashberg theory
\cite{Scalapino}, and deriving $2\times 2$ self-energy matrices
$\underline{\Sigma}^q({\bf k},z) = \Sigma^q({\bf k},z)
\underline{\tau}_3 + \Re \Phi^q({\bf k},z) \underline{\tau}_1 + \Im
\Phi^q({\bf k},z) \underline{\tau}_2$, where $\underline{\tau}_1$,
$\underline{\tau}_2$ and $\underline{\tau}_3$ are the Pauli matrices,
and $\Phi^q$ is the pairing order parameter. The diagonalization of
$\epsilon^q_0({\bf k}) \underline{\tau}_3 + \underline{\Sigma}^q({\bf
k},z)$ yields Bogoliubov states created by $q_{+}^{\dagger}({\bf k},z)$,
and $q_{-}^{\dagger}({\bf k},z)$. Dyson's equation yields QE GF poles at
$z = \pm \epsilon^q_0({\bf k}) + \Sigma_{\pm}^q({\bf k},z) =
E_{\pm}^q({\bf k},z) + i\cos{(2\xi_{\bf k}^q(z))} \Im \Sigma^q({\bf
k},z)$, where $E_{\pm}^q({\bf k},z) = \pm \{ [\epsilon^q_0({\bf k}) +
\Re \Sigma^q({\bf k},z)]^2 + [\Re \Phi^q({\bf k},z)]^2\}^{\half}$, and
$\sin{(2\xi_{\bf k}^q(z))} = \Re \Phi^q({\bf k},z) / E_{+}^q({\bf
k},z)$. The spectral functions $A_{\pm}^q({\bf k},\omega)$ are derived
through an equivalent expression to Eq.~(\ref{eq8}), replacing
$\epsilon^q_0({\bf k}) + \Sigma^q({\bf k},\omega - i0^+)$ by $\pm
\epsilon^q_0({\bf k}) + \Sigma_{\pm}^q({\bf k},\omega - i0^+)$.
$\Sigma^q({\bf k},z)$ and $\Phi^q({\bf k},z)$ are evaluated on the basis
of similar expressions to Eq.~(\ref{eq7}), where $A^q({\bf k}-{\bf q},
\omega^q)$ is replaced by $\cos{^2(\xi_{{\bf k}-{\bf q}}^q(\omega^q))}
A_{+}^q({\bf k}-{\bf q}, \omega^q) + \sin{^2(\xi_{{\bf k}-{\bf
q}}^q(\omega^q))} A_{-}^q({\bf k}-{\bf q}, \omega^q)$, for $\Sigma^q$,
and by $\half \sin{(2\xi_{{\bf k}-{\bf q}}^q(\omega^q))} [A_{+}^q({\bf
k}-{\bf q}, \omega^q) - A_{-}^q({\bf k}-{\bf q}, \omega^q)]$, for
$\Phi^q$. 

The derivation of low-energy  $\Sigma^q$ and $\Phi^q$ is expressed in
terms of contributions due to coupling to peaks of QE-lagron states of
energies $\sim \omega^q \pm \omega^{\lambda}({\bf Q}_m) \simeq
\omega^q$, at points $\sim {\bf k} + {\bf Q}_m$, and of energies $\sim
\omega^q + \text{sign}(\omega^q) \omega^{\lambda}({\bf Q})$ at points
$\sim {\bf k} + {\bf Q}$. The resulting nonzero $\Re \Phi^q({\bf
k},\omega)$ does not change its sign as a function of $\omega$, and
reverses it (maintaining its magnitude) when ${\bf k}$ is shifted by
${\bf Q}$, consistently with an approximate $d_{x^2-y^2}$ pairing. The
coupling of stripons and arcons to the above ({\it both} positive- and
negative-energy) peaks helps stabilize a nonzero $\Phi^q$, while terms
of opposite signs are introduced to $\Sigma^q$. Consequently, the effect
of $\Phi^q$ on the gap dominates. 

Since QE coupling through ${\bf Q}_m$ lagrons is maximal for stripons,
pairing is strongest for them (note that unlike the coupling constants
between electrons and acoustic phonons which vanish for ${\bf q} \to 0$,
$\gamma({\bf q})$ remain constant for ${\bf q} \to {\bf Q}_m$).
Consequently, there exists a temperature range $T_c < T < T^*$, where
$\Phi^q \ne 0$ for stripons and humpons, but {\it not} for arcons, and
it corresponds to the pseudogap (PG) phase \cite{Honma}. The stripon
peaks then split into positive- and negative-energy ones, as is sketched
in  Fig.~\ref{fig2}(a), and their signature on the electron spectrum in
the PG phase, is sketched in Fig.~\ref{fig2}(b). The opening of a
stripon gap causes a large reduction of the scattering rates below those
of the MFL phase for $\omega$ below the gap energy, and much less above
it, in agreement with experiment \cite{Takagi, Puchkov}. A decrease in
$T$ ($>T_c)$ results in the transition of arcons into stripon-humpons
which would include all the arcons if $\delta {\bf q}_m$ were small
enough (see Fig.~\ref{fig1}(c)). In such a case, the extension of $T^*$
($>T_c$) to zero would result in the reduction of the Fermi arcs into
points \cite{Kanigel}. 

SC occurs when $\Phi^q\ne 0$ {\it also} for arcons which are not coupled
to other QE's through ${\bf Q}_m$ lagrons, and it is induced through the
${\bf Q}$-ESP lagrons. Since the energy difference involved is $\sim
\omega^{\lambda}({\bf Q}) \simeq E_{_{\rm RM}}$ (see above), it scales
with $k_{_{\rm B}}T_c$, with a factor $\sim 5$ (see
Ref.~\cite{AshkHam}), similarly to the ratio between the energy of the
relevant phonons and $k_{_{\rm B}}T_c$ in strong-coupling
phonon-mediated SCs. Thus the arcon peaks which are not on the lines of
nodes split below $T_c$, resulting in the modification of the gap
structure which resembles \cite{Shen2} the ``onset of a second energy
gap'', scaling with $\omega^{\lambda}({\bf Q}) \simeq E_{_{\rm RM}}
\simeq 5 k_{_{\rm B}}T_c$. 

The opening of an SC gap diminishes low-energy scattering of gap-edge
QE states, resulting in sharp split stripon and arcon peaks below $T_c$,
while the humpon peaks remain wide (see Fig.~\ref{fig2}(a)). The
signature of the split peaks on the electron spectrum in the SC phase is
sketched in Fig.~\ref{fig2}(b). Since the SC gap is almost complete, the
scattering rates are radically reduced below $T_c$ \cite{Puchkov} for
$\omega$ below the gap energy, while the effect above it is small. This
results also in small-linewidth low-energy svivons, and thus
\cite{AshkHam} in a sharp RM. Since (see above) the energy separation
between the stripon and the humpon $>\omega^{\lambda}({\bf Q})$, a dip
appears \cite{Zasadzinski} between the peak and the hump, at about
$\omega^{\lambda}({\bf Q}) \simeq E_{_{\rm RM}}$ above it. 

The values of $\omega^{\lambda}({\bf q})$ and $\gamma({\bf q})$ are
determined through the condition that the same
``constraint-susceptibility'' (CS) \cite{AshkHam} is obtained using
either the svivon or the QE spectrum. A major feature of the CS, based
on the svivon spectrum in the SC state, is \cite{AshkHam} the existence
of a peak around ${\bf k}=0$ at an energy $\sim E_{_{\rm RM}}$. When it
is calculated using the QE spectrum, this CS peak represents some
average of the QE SC gap, and contributions to it cancel out
\cite{AshkHam} at points where $\cos{(2\xi_{\bf k}^q(\omega^q))} =
[\epsilon^q_0({\bf k}) + \Re \Sigma^q({\bf k},\omega^q)] / E_{+}^q({\bf
k},\omega^q) = 0$. For stripons and arcons one has $\epsilon^q_0({\bf
k}) + \Re \Sigma^q({\bf k},\omega^q) \simeq 0$ for $\Phi^q= 0$ but {\it
not} for $\Phi^q\ne 0$. In the case of stripons, the modification of
$\Sigma^q({\bf k},\omega^q)$ for $\Phi^q\ne 0$ includes contributions of
opposite signs, due to coupling to positive- and negative-energy
stripons through ${\bf Q}_m$ lagrons. Thus, fine tuning of
$\omega^{\lambda}({\bf q})$ and $\gamma({\bf q})$ yields smaller
$\cos{(2\xi_{\bf k}^q(\omega^q))}$ values for stripons than for arcons,
such that the low-energy CS peak around ${\bf k}=0$ corresponds to $\sim
E_{_{\rm RM}}$, as is required. 
 
In conclusion, a unified theory for the cuprates and related SCs, where
the low-energy excitations are approached in terms of combinations of
atomic-like electron configurations, and a Lagrange Bose field enables
treating them as bosons or fermions, has been shown to resolve mysteries
of the cuprates. This includes their observed phase diagram, non-FL to
FL crossover, the existence of MFL behavior and a PG phase with Fermi
arcs, kink- and waterfall-like spectral features, the drop in the
scattering rates in the PG phase, and further in the SC phase, an
effective increase in the density of carriers with $T$ and $\omega$, the
correspondence between $T_c$, $E_{_{\rm RM}}$, and the increase in the
gap below $T_c$, {\it etc}. Similar anomalies to those of the cuprates
exist in the FeSCs, and a forthcoming paper will address {\it both}
similarities and differences between them. 



\begin{thebibliography}{99}
\bibitem{Takagi}H.~Takagi, {\it et al}, {\it Phys.~Rev.~Lett.} {\bf 69},
2975 (1992). 
\bibitem{Hwang}H.~Y.~Hwang, {\it et al}, {\it Phys.~Rev.~Lett.} {\bf
72}, 2636 (1994). 
\bibitem{Fisher}J.~Genossar, {\it et al}, {\it Physica C} {\bf 157}, 320
(1989). 
\bibitem{Tanaka}S.~D.~Obertelli, {\it et al}, {\it Phys.~Rev.~B} {\bf
46}, 14928 (1992). 
\bibitem{Puchkov}A.~V.~Puchkov, {\it et al}, {\it J.~Phys.: Cond.~Mat.},
{\bf 8}, 10049 (1996). 
\bibitem{Zasadzinski}J.~F.~Zasadzinski, {\it et al}, {\it
Phys.~Rev.~Lett.} {\bf 87}, 067005 (2001). 
\bibitem{Luo}H.~G.~Luo, {\it et al}, {\it Phys. Rev. B} {\bf 77}, 014529
(2008). 
\bibitem{Tanner1}D.~B.~Tanner, {\it et al}, {\it Physica B} {\bf 244}, 1
(1998). 
\bibitem{Boeb}G.~S.~Boebinger, {\it et al}, {\it Phys.~Rev.~Lett.} {\bf
77}, 5417 (1996). 
\bibitem{Anderson}P.~W.~Anderson, {\it Phys.~Rev.~Lett.} {\bf 64}, 1839
(1990). 
\bibitem{Varma}C.~M.~Varma, {\it et al}, {\it Phys.~Rev.~Lett.} {\bf
63}, 1996 (1989). 
\bibitem{AshkHam}J.~Ashkenazi, arXiv:0809.4237, {\it J. Supercond. Nov. 
Magn.} DOI: 10.1007/s10948--008--0370--8.
\bibitem{Barnes} S.~E.~Barnes, {\it Adv.~Phys.} {\bf 30}, 801-938
(1981). 
\bibitem{Honma}T.~Honma, and P.~H.~Hor, {\it Phys.~Rev.~B} {\bf 77},
184520 (2008). 
\bibitem{Campuzano}K.~Nakayama, {\it et al}, {\it Phys.~Rev.~B} {\bf
74}, 054505 (2006). 
\bibitem{Chang}J.~Chang, , {\it et al}, {\it New J. Phys.} {\bf 10},
103016 (2008). 
\bibitem{Shen1}W.~Meevasana, {\it et al}, {\it Phys.~Rev.~B} {\bf 77},
104506 (2008). 
\bibitem{Scalapino}D.~J.~Scalapino, , {\it et al}, {\it Phys.~Rev.} {\bf
148}, 263 (1966). 
\bibitem{Kanigel}A.~Kanigel, {\it et al}, {\it Nature Physics} {\bf 2},
447 (2006). 
\bibitem{Shen2}W.~S.~Lee, {\it et al}, {\it Nature} {\bf 450}, 81
(2007). 
\end{thebibliography}
\end{document}